# The Death of Renaissance Scientist

Zak Risha[1], Yiling Lin[1], Erin Leahey[2*] and Lingfei Wu[1*]

[1]School of Computing and Information, The University of Pittsburgh. [2] School of Sociology, University of Arizona.

*Corresponding author. E-mail: liw105@pitt.edu (L.W) and leahey@arizona.edu (E. L)

**Abstract.** Scholars are often categorized into two types: hedgehogs (specialists), who focus on working within a specific research field, and foxes (generalists), who actively contribute to a variety of fields. Despite the familiar anecdotes and popularity of this distinction, its empirical foundation has remained largely unexamined. We examine whether the research style of being a fox or a hedgehog is a stable personal trait or an evolving strategy over a scientist's career. Analyzing 2.3 million scholars' publication records over a century, we find that research styles exhibit remarkable stability. Notably, the proportion of fox-like scientists has dramatically declined in the past century, a phenomenon we term "the death of Renaissance scientists." This decline is particularly significant as science shifts toward team collaboration. Teams of foxes consistently outperform teams of hedgehogs in generating new ideas and directions, as confirmed by two emerging innovation metrics for papers: atypicality and disruption. Our research is the first to quantify the process and consequences of the decline of Renaissance scientists. By doing so, we establish a universal link between research styles, demographic shifts, and innovative output.

**Introduction**

In the past century, the era of the lone genius in science has given way to the teamwork of ordinary people (1). This historical trend, referred to as "the death of Renaissance Men," is accompanied by the increasing complexity of scientific problems—from global warming to new, easily transmissible diseases—and the narrowing of individual scientists' expertise (2). In this context, the idea of assembling specialized researchers to pursue groundbreaking discoveries, akin to gathering a blacksmith, a mason, and a carpenter to mirror Leonardo da Vinci's genius, has gained momentum and fueled the proliferation of teams across various scientific domains, especially interdisciplinary ones (3, 4).

However, despite the global increase in team-based interdisciplinary research, science has not produced the expected surge in innovative work. Recent research reveals a decline in scientific breakthroughs (5), documents that adding more experts to a team does not always enhance innovation (6), and challenges the effectiveness of the division of labor in conceptual work (7, 8). This raises fundamental questions: Is the era of Renaissance Men truly ending, and if so, what are the implications for innovation?

To investigate, we analyze the publication records of 29 million name-disambiguated scholars over a century. Following philosopher Isaiah Berlin's classification (9), we distinguish between two types of scholars: hedgehogs (specialists), who focus on a specific research field, and foxes (generalists), who contribute to a wide variety of fields. To do so, we propose an intuitive metric, the Specialization (S) Index, to measure a scientist's research concentration by their maximum proportion of publications within a single field. We find that S-Index exhibits remarkable stability over the course of an academic career, suggesting that research style is a personal trait rather than a shifting strategy, which contrasts with previous studies (10, 11). Additionally, we find that the proportion of fox-like scientists has dramatically declined over the past century, confirming the "death of Renaissance Men" assumption (2). Finally, we examine an entirely fresh aspect: the impact of this decline on team innovation. Our analysis demonstrates that teams of foxes consistently outperform hedgehogs in generating new ideas, as evidenced by two emerging innovation metrics for papers, disruption (5, 6) and atypicality (12).

**Results**

We began by identifying scientists' research styles from their published articles. Using an open-access, validated scientific taxonomy proposed by Microsoft Academic Graph (13) and incorporated into OpenAlex (14), we categorize research articles into 292 research fields, including Discrete Mathematics, Molecular Biology, and Organic Chemistry (15). This classification offers similar granularity to previous measures like the Physics and Astronomy Classification Scheme (PACS) (11, 16) and Mathematics Subject Classification (MSC) (10) but provides much broader coverage across the full spectrum of science. On average, scientists publish in four fields across their careers. We quantify a scientist's research concentration using the maximum proportion of their publications within a single field, called the specialization (S) score. A scientist with an S-score of 0.5 or greater is classified as a "hedgehog" (specialist), as more than half of their papers are concentrated in their primary research field. Conversely, a scientist with an S-score below 0.5 is classified as a "fox" (generalist), as their papers are distributed across various fields (Figure 1a).

Next, we examined whether the research style of being a fox or a hedgehog is a stable personal trait or a shifting strategy over an individual's career (Figure 1b). To do this, we followed previous studies in analyzing scientists' personal traits (17) and examined the temporal change of S-scores against the accumulated fraction of papers throughout their careers. We found that most scientists (82%) consistently stay above or below 0.5, indicating that such changes do not typically shift research styles. These results suggest that research style is a relatively stable personal trait (16, 18), challenging previous research that assumed it to be a shifting strategy over academic careers without robust testing (10, 11).

Using these consistent research styles, we observed a dramatic decline in the proportion of fox-like scientists, dropping by 15% over half a century, from approximately 55% in the 1960s to 40% in the 2010s (Figure 1c). Based on previous literature, which conjectured but did not empirically verify this pattern (2), we term this phenomenon "the death of Renaissance scientist."

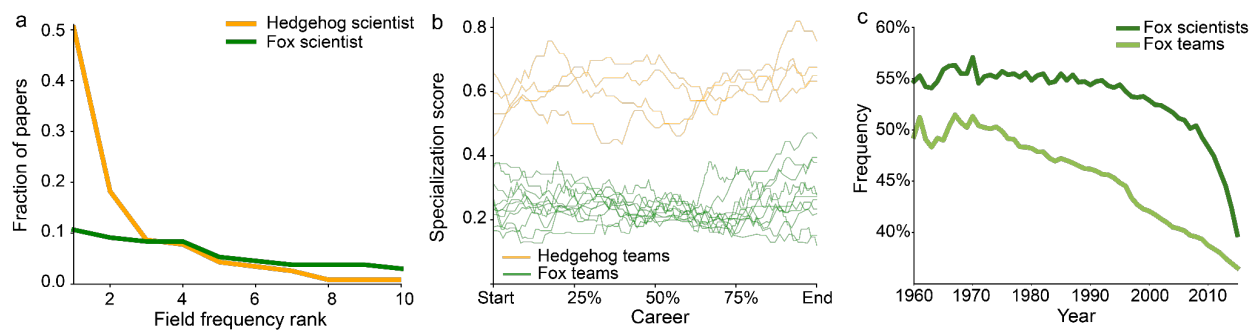

**Figure 1. The Research Style of Scientists and The Death of Renaissance Men.** We analyzed the research styles of scientists across 29,690,219 journal articles from 1960 to 2020. We focused on 2,301,341 scientists who authored

three or more papers before 2015 to ensure robust estimations of their research style and impact. Using their first-authored works, we calculated their specialization (S) score as the fraction of papers in their most frequent field. Panel (a) shows an example of a hedgehog-like scientist (S=0.5, in orange) and a fox-like scientist (S=0.1, in green). For 432,085 scientists with ten or more papers, we calculated their S-score using a moving average across a window of 10 papers. Most scientists (82%) remained consistently above or below 0.5. For clarity, Panel (b) plots the trends for scientists with 100 or more papers, using a moving average across a window of 30 papers, and with S-scores around 0.2 (fox) and 0.6 (hedgehog). Panel (c) illustrates a decline in fox-like scientists from 55% to 40% (dark green) and in teams of only fox-like scientists from 50% to 36% (light green).

Identifying research styles enabled us to distinguish between teams comprising generalists and those of specialists and compare their performance in interdisciplinary teams (Figure 2a). We focus on interdisciplinary teams for two reasons. First, these teams are designed to integrate diverse knowledge, allowing us to compare generalists (who possess interdisciplinary knowledge individually) with specialists from different fields (who represent interdisciplinary knowledge as a group) (19). In other words, we aim to understand the effect of research style while controlling for knowledge diversity. To simplify this comparison, we excluded teams that mix specialists with generalists. Second, this comparison has strong policy implications, reflecting the classic dilemma research institutes face when forming interdisciplinary groups: should they hire foxes or hedgehogs (20)?

To compare the innovative performance of teams, we calculated two key innovation metrics. The Disruption (D) index measures idea displacement (21), quantifying how subsequent papers cite a focal paper while disregarding its references (5, 6). The Atypicality (A) index measures idea recombination, assessing a paper's novelty by quantifying how it unexpectedly cites and combines prior work from different scientific journals (12). In our analysis, we propose inverting the negative z-score used in (12), so that higher positive values signify greater atypicality, reflecting advancements in discovering surprising and complementary knowledge.

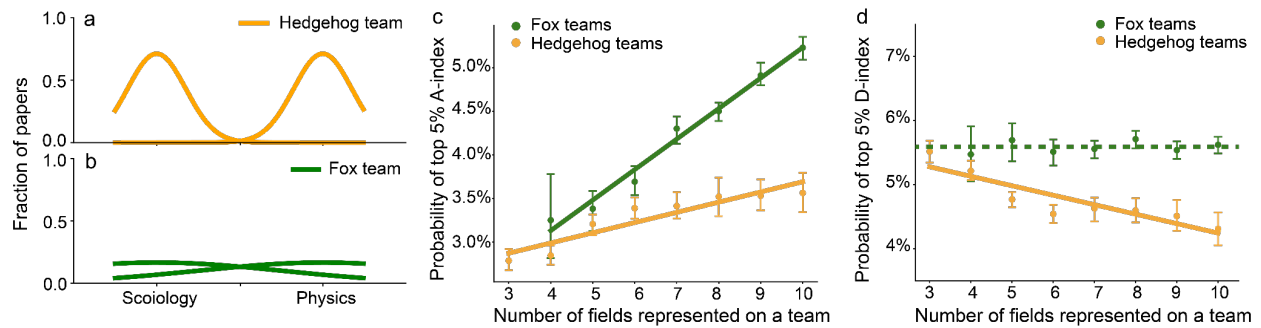

**Figure 2. Fox Teams Are More Innovative Than Hedgehog Teams.** We analyzed 4,135,914 papers by 2,301,341 scientists, distinguishing hedgehog teams (specialists) from fox teams (generalists). In Panels (c-d), we compared their innovation performance. Panel (c) uses the Atypicality (A) index to measure idea recombination, assessing novelty by how unexpectedly papers cite and combine prior work from different journals. Fox teams (green line) increased the combination of surprising ideas faster than hedgehog teams (orange line) as the team incorporated more fields. Panel (d) uses the Disruption (D) index to measure idea displacement, assessing how subsequent papers cite a focal paper while disregarding its references. Fox teams' (green line) likelihood of producing highly disruptive work remained stable even with more fields. In contrast, hedgehog teams' (orange line) probability of disruption declined by 22% (from 5.5% to 4.3%) as they integrated additional fields. Error bars in (b and c) indicate a 95% bootstrap confidence interval centered at the mean.

We found that generalist teams consistently outperform specialist teams across both innovation metrics. As the number of knowledge domains added to a team grows, both generalist and specialist teams increase their ability to produce highly atypical work, but generalists increase faster (Figure 2c). Moreover, as they integrate additional knowledge domains, generalist teams maintain their ability to produce highly disruptive work, while specialist teams do not (Figure 2d). For instance, when the number of fields increases from three to ten, specialist teams' likelihood of producing highly disruptive work declines by 22% (from 5.5% to 4.3%). The advantage of generalist teams holds even after controlling for team size, publication year, and scientists' average career age (see SM).

**Discussion**

Our results suggest that even as scientists become more specialized, simply assembling specialists from different fields may not be enough to foster breakthrough ideas. Instead, successful and innovative collaborations that truly alter knowledge flows are more likely when teams are composed of generalists—scientists who integrate interdisciplinary links within a single brain (22). While their fields of expertise may not align perfectly (only 2% of generalist teams possess entirely overlapping expertise among team members), these broad thinkers share a research style that, when combined, nurtures intellectual synergies conducive to the production of highly disruptive work. They are likely better equipped than teams of specialists to address the epistemological and communicative challenges of interdisciplinary collaboration (4, 23), and to realize the expected impact of such collaboration. This approach's success is exemplified by the partnership between Herbert A. Simon and Allen Newell, both fox-like scientists in our dataset, who pioneered the field of artificial intelligence.

These findings highlight the complexity of scientific discovery, which resists being broken into discrete tasks across team members (24). As knowledge burdens increase, scientists often form teams with complementary specialties, dividing tasks to extend their reach (3). External funding agencies support this by promoting cross-cutting programs and large, diverse teams for convergent and translational research (20, 25). However, our results suggest that science policymakers should reconsider this approach. To address grand challenges and spur innovation, it may be more effective to foster "intra-personal interdisciplinarity" (19) in individual scientists and form teams of broad-thinking generalists.

**Materials and Methods**

**Datasets.**

Our research is based on the Microsoft Academic Graph (MAG), which provides a name-disambiguated database comprising 28 million scientists and 29 million journal articles between 1960 and 2020 (13). We selected scientists who started publishing prior to 2015, which guarantees that each scientist has accumulated publications spanning at least five years, ensuring a robust estimation of their research impact. To classify scientists' work into fields, we rely on an established and validated scientific taxonomy of 292 research fields, including Discrete Mathematics, Molecular Biology, and Organic Chemistry (13).

# Supplementary Materials

To eliminate the influence of potential confounding factors that might account for the superior performance of generalist teams over specialist teams, we specified multivariate regression models. The outcome of interest is binary – whether a paper falls in the top 5th percentile of the Disruption Index distribution – so we use Logistic Regression. We control for team size, publication year, and the average career age of team members. We find that relative to specialist teams, generalist teams are 31% more likely to produce highly disruptive papers (see Table 1). Moreover, judging from the odds ratio, this effect is quite large relative to the effect of other variables like team size, year of publication, and average career age. This finding was corroborated when we measured scientists' research styles at the discipline level (using 19 categories rather than 292 fields). Results from these multivariate models indicate that, while relevant to disruptive potential, these other factors do not account for the effect of generalist teams that we document.

Table 1. Regression models of probability to disrupt

|  | 292 Fields | | 19 Disciplines | |
| --- | --- | --- | --- | --- |
|  | Coef | Odds Ratio | Coef | Odds Ratio |
| Is Generalist Team | .267*** | 1.31 | .608*** | 1.84 |
| Team size | -.214*** | 0.81 | -.186*** | 0.83 |
| Year of publication | -.053*** | 0.95 | -.056*** | 0.95 |
| Average career age | -.004*** | 0.99 | -.006*** | 0.99 |
| n | 2163661 | | 2916089 | |